\documentclass[twocolumn,twoside,slac_two]{revtex4}
\usepackage{graphicx}
\usepackage{fancyhdr}
\begin{document}

\newcommand{\farcm}{\mbox{\ensuremath{.\!\!^{m}}}}
\newcommand{\fm}{\mbox{\ensuremath{.\!\!^{m}}}}
\newcommand{\farcs}{\mbox{\ensuremath{.\!\!^{\prime\prime}}}}
\newcommand{\fs}{\mbox{\ensuremath{.\!\!^{s}}}}

\title{Detection of High-Energy Gamma-Ray Emission from the Globular Cluster 
47 Tucanae with Fermi}

%

\author{N.A. Webb, J. Kn\"odlseder}
\affiliation{Universit\'e de Toulouse, UPS, CESR, 9 Avenue du Colonel Roche, F-31028 Toulouse Cedex 9, France \\ on behalf of the Fermi Large Area Telescope Collaboration}
%

\begin{abstract}

Globular clusters are known to harbour a significant population of
neutron star X-ray binaries that could be responsible for delaying the
inevitable core collapse of these dense clusters.  As a result, their
progeny, namely millisecond pulsars, are also present in large numbers.
Following the confirmation using the Fermi Gamma-ray Space Telescope
that millisecond pulsars are indeed gamma-ray emitters, we report on
the detection of the Galactic globular cluster 47 Tuc with the Fermi
Large Area Telescope.  This is the first detection of a Galactic
globular cluster in the gamma-ray domain.  The gamma-ray spectrum is
consistent with gamma-ray emission from a population of millisecond
pulsars. The observed gamma-ray luminosity implies an upper limit of
60 millisecond pulsars present in 47 Tucanae. 

\end{abstract}

\maketitle

\thispagestyle{fancy}


\section{INTRODUCTION}
With their typical ages of $\sim$10$^{10}$ years, globular
clusters form the most ancient constituents
of our Galaxy. They are seen
throughout the electromagnetic spectrum, from radio
waves to X-ray energies, revealing their various
stellar components. As an example, X-ray observations
have shown that globular clusters contain considerably
more close binary systems per unit mass than
the Galactic disc [1]; this finding is interpreted as a
result of frequent stellar encounters in their dense
stellar cores [2]. This scenario is strengthened by the
observation that the number of low-mass X-ray binary
systems containing neutron stars is directly correlated
with the stellar encounter rate [3, 4]. These
close binary systems may provide a source of internal
energy stabilizing the cluster against the inevitable core collapse [5].
Another consequence of this scenario is the presence
of many millisecond pulsars (MSPs, also known as
{\em recycled} pulsars); these are pulsars that were spun 
up to millisecond periods by mass accretion from a
low-mass X-ray binary companion [6].

The only domain in which globular clusters have
so far eluded detection is gamma rays. Recent observations
with the Large Area Telescope (LAT) onboard
the Fermi Gamma-ray Space Telescope have
revealed gamma-ray pulsations from eight MSPs,
establishing these objects as a class of high-energy
gamma-ray sources [7, 8]. Most of the MSPs detected
in gamma rays are within a distance of only a
few hundred parsecs of the Sun, which implies
that MSPs are rather faint objects with isotropic
gamma-ray luminosities that generally do not exceed
10$^{33}$ ergs s$^{-1}$ [8]. Placed at a distance of a few kiloparsecs
(the distance to the nearest globular clusters),
it is unlikely, although not impossible, that individual
MSPs are being detected in gamma rays. Globular
clusters, however, may contain tens to several hundreds
of MSPs [9], and their cumulative magnetospheric emission is probably the first signature
that would be picked out in gamma rays.

47 Tucanae (NGC 104) is one of the most promising
candidates for high-energy gamma-ray emission
because of the large number of known MSPs
in the cluster and its relative proximity (4 kpc) [10].
So far, 23 MSPs have been detected in 47 Tuc through
radio and/or X-ray observations, and the total population
is estimated to be between 30 and 60 [9, 11, 12], although
claims in the past reached up to 200 [13].

\section{DATA ANALYSIS}

We have observed 47 Tuc with the LAT telescope aboard Fermi. Our data
amount to 194.3 days of continuous sky survey observations over the
period August 8th 2008 - April 3rd 2009 during which a total exposure
of $\sim$2 $\times$ 10$^{10}$ cm$^2$ s (at 1 GeV) has been obtained for 47 Tuc. Events
for the data taking period satisfying the standard low-background
event selection ({\em Diffuse} events [14]) and coming from zenith angles
$<$ 105$^\circ$ (to greatly reduce the contribution from Earth albedo gamma
rays) were used. To further reduce the effect of Earth albedo
backgrounds, the time intervals when the Earth was appreciably within
the field of view (specifically, when the center of the field of view
was more than 47$^\circ$ from the zenith) were excluded from this analysis,
and all events taken when the spacecraft was within the South Atlantic
Anomaly were also excluded.  The data analysis presented in this paper
has been performed using the LAT Science Tools package, which is
available from the Fermi Science Support Center, using P6 V3
post-launch instrument response functions (IRFs). These take into
account pile-up and accidental coincidence effects in the detector
subsystems that are not considered in the definition of the pre-
launch IRFs.  Using a maximum likelihood model fitting procedure we
determine the position of the gamma-ray source to be ($\alpha_{2000}$ = 0$^h$24$\fm$3, $\delta_{2000}$ = -72$^\circ$03$\farcm$8) with a 95\% confidence error radius of
4.2'. Systematic uncertainties in the position due to inaccuracies in
the point-spread function and the telescope alignment are estimated to
be $<$ 1'. The position of the gamma-ray source is spatially consistent
with the location ($\alpha_{2000}$ = 0$^h$24$^m$05$\fs$67, $\delta_{2000}$ = -72$^\circ$04'52$\farcs$62) of
the core of 47 Tuc [15]. We tested for a possible extent of the
gamma-ray emission by fitting 2D Gaussian-shaped intensity profiles to
the LAT data for which we adjusted the widths and positions. We find
that the emission is best fitted as a point source. From the decrease
of the likelihood with increasing Gaussian width we derive an upper
limit for the extent (FWHM) of 21' (2$\sigma$). This limit is considerably
larger than the 47 Tuc core radius of 25'' [16], demonstrating that
47 Tuc's core cannot be resolved by the LAT.

\begin{figure}
\includegraphics[width=80mm]{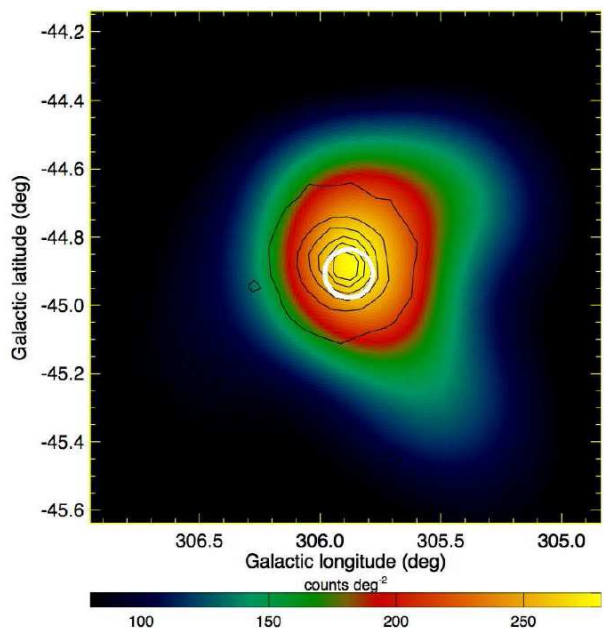}
\caption{Fermi LAT gamma-ray image (200 MeV to 10 GeV) of a 1.5$^\circ$ $\times$ 1.5$^\circ$ region centered on the position
of 47 Tuc. The map was adaptively smoothed by imposing a minimum signal-to-noise ratio of 5. A total of $\sim$290 counts were detected from the gamma-ray source. Black contours indicate the stellar density in 47 Tuc
as derived from DSS2 red plates [32]. The white circle shows the 95\% confidence region for the location of
the gamma-ray source. The position of the LAT source coincides with the core region of 47 Tuc.}
\label{fig:47tucimage}
\end{figure}

We determined the gamma-ray spectrum of the LAT
source by maximum likelihood fitting of the emission in 10
logarithmically spaced energy bins covering the interval 200 MeV to 10
GeV (Fig. 2). The spectrum reveals a relatively flat spectral energy
distribution with a clear cutoff at energies above a few GeV; it is
well fitted by an exponentially cut-off power law that provides a 3$\sigma$
improvement upon a simple power law, with best-fitting spectral index $\Gamma$
= 1.3 $\pm$ 0.3 and cutoff energy E$_{cut}$ = 2.5$\pm^{\scriptscriptstyle 1.6}_{\scriptscriptstyle 0.8}$ GeV. The systematic
uncertainty in the spectral index is estimated to be 0.1; that in the
cut-off energy is estimated to be 0.3 GeV.  By integrating the
best-fitting model over the energy range 100 MeV to 10 GeV, we
determined the integral photon flux in this band to be 2.6 ($\pm$0.8) $\times$
10$^8$ photons cm$^{-2}$ s$^{-1}$, which is slightly below the upper limit of 5 $\times$
10$^8$ photons cm$^{-2}$ s$^{-1}$ reported by EGRET [17-19].  The photon flux
corresponds to an energy flux of 2.5 ($\pm$0.4) $\times$10$^{-11}$ ergs cm$^{-2}$ s$^{-1}$.  The
systematic uncertainty in our fluxes is estimated to be $<$10\%. The
overall detection significance of the source amounts to 17$\sigma$.  

We
searched for time variability of the gamma-ray signal by dividing our
data set into equally sized time bins of durations 1 day, 1 week, 2
weeks, and 1 month.  We detected no source at the location of 47 Tuc in
any of the daily or weekly time bins, which indicates that the
observed emission did not arise from short-duration flares. The LAT
source was significantly detected in all 2-week and monthly time bins
at a comparable flux level, which suggests that the source was steady
over the period of observation.  Using ephemerides from [18] for
21 MSPs in 47 Tuc, we searched for gamma-ray pulsations in our data
without finding any significant detection. The observed gamma-ray
signal thus does not appear to be dominated by a single (or a few)
known gamma-ray pulsars in 47 Tuc; this is in line with the absence of
a single particularly powerful MSP in the cluster.

\begin{figure}
\includegraphics[width=80mm]{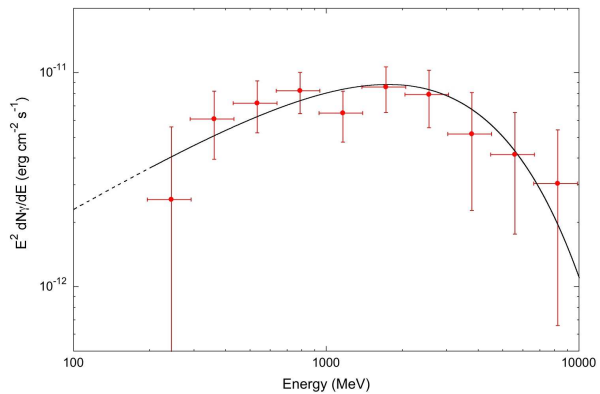}
\caption{Spectral energy distribution
(E$^2$dN$_\gamma$/dE) of the
Fermi source seen toward
47 Tuc. The solid line shows
the fit of an exponentially
cut-off power law obtained
for the energy range 200 MeV
to 20 GeV. The dashed line
indicates the extrapolation of
the fit to 100 MeV.}
\label{fig:47tucspectrum}
\end{figure}

\section{DISCUSSION}

Pulsed gamma-ray curvature radiation (and eventually inverse Compton
scattering) arising near the polar cap and/or in an outer
magnetospheric gap in MSPs has been proposed as a possible source of
high-energy photons in globular clusters [21-24].  The main unknowns of
this scenario are the exact site of gamma-ray production (polar cap
versus slot gap or outer gap), the efficiency $\eta_\gamma$ with which the
spin-down power is converted into gamma-ray luminosity, and the total
number of MSPs in the cluster. The eight Galactic MSPs that have so far
been detected by Fermi [8] have a mean spectral index of $< \Gamma >$ = 1.5 $\pm$ 0.4 and a mean cutoff energy of $< E_{cut} >$ = 2.8 $\pm$ 1.8 GeV, similar
to that which we found for the 47 Tuc source.  Cumulative gamma-ray emission
from MSPs in 47 Tuc is thus a plausible explanation for the observed
signal.  

Under this hypothesis, we estimate $\eta_\gamma$ from the gamma-ray flux
of 47 Tuc by taking the average spin-down power to be $< $ \.E $ >$ = 1.8
($\pm$0.7) $\times$ 10$^{34}$ ergs s$^{-1}$.  Keeping the total number of MSPs in 47
Tuc as a parameter of the solution, for a distance to 47 Tuc of 4.0 $\pm$
0.4 kpc [10], the measured energy flux of 2.5 ($\pm$0.4) $\times$ 10$^{-11}$ ergs cm$^{-2}$
s$^{-1}$ converts into an isotropic gamma-ray luminosity of L$_\gamma$ = 4.8 ($\pm$ 1.2) $\times$ 10$^{34}$ ergs
s$^{-1}$.  This results in $\eta_\gamma$ = 0.12 ($\pm$0.05) $\bar{f}_\Omega$ / N$_{23}$, where $\bar{f}_\Omega$
is an average geometrical correction factor that accounts for
non-isotropic emission [25] and N$_{23}$ is the number of MSPs in 47 Tuc in
units of 23. N$_{23} \geq $ 1 implies that 0.12 ($\pm$0.05) $\bar{f}_\Omega$ is an upper limit
on the spin-down to gamma-ray luminosity conversion efficiency in 47
Tuc; this is consistent with the predicted conversion efficiency of
6.1\% based on a model of the expected gamma-ray emission from a
population of MSPs in 47 Tuc in the framework of a fully
three-dimensional general-relativistic polar cap pulsar model [24]. It
is also compatible with the estimated conversion efficiency of $\sim$10\% of
[26] that was derived in the framework of a space charge-limited flow
acceleration polar cap model.  

The conversion efficiencies of the
eight Galactic MSPs detected by Fermi cover the range from 0.02$\bar{f}_\Omega$  to
1.0$\bar{f}_\Omega$  with a mean value of $<$ $\eta_\gamma >$ = 0.14 $\bar{f}_\Omega$ [8], which is larger than
the upper limit we derive for 47 Tuc. However, the MSPs that have been
detected so far by Fermi are faint gamma-ray sources, forming a sample
that is likely biased toward intrinsically bright objects or objects for which the beam orientation is
favorable. Selecting only the nearest MSPs should
reduce this bias, because for close objects a larger
fraction of the MSP parameter space is accessible to
Fermi. Taking only the three nearest MSPs from the
Fermi sample (which also corresponds to the three
nearest known MSPs) results in a mean spin-down
to gamma-ray luminosity conversion efficiency of
$<$ $\eta_\gamma >$ = 0.08 $\bar{f}_\Omega$ , considerably lower than the global
average. Taking the five nearest MSPs results in
$<$ $\eta_\gamma >$ = 0.10 $\bar{f}_\Omega$ . Both values are consistent with our
upper limit on 47 Tuc. Thus, our data show no evidence
for differences in the gamma-ray efficiencies of
MSPs in globular clusters with respect to objects
observed in the Galactic field.

Assuming that the gamma-ray efficiencies of
MSPs in 47 Tuc are equal to those of the nearby
Galactic field sample, and also assuming that their
average geometrical correction factors $\bar{f}_\Omega$ are the
same, we obtained an estimate of the total number of
MSPs in 47 Tuc.  Taking the mean  $<$ $\eta_\gamma >$ = 0.08 $\bar{f}_\Omega$ that
we obtained for the sample of the nearest Galactic
MSP as the most conservative estimate, we converted
the observed gamma-ray efficiency $\eta_\gamma$ = 0.12 ($\pm$0.05) $\bar{f}/N_{23}$ for 47 Tuc into N$_{23}$ = 1.5 $\pm$ 0.6. Formally,
this corresponds to a 95\% confidence interval of 7 to
62 MSPs in 47 Tuc (assuming that uncertainties are
Gaussian), which is consistent with the range of 45 to
60 MSPs estimated on the basis of X-ray observations
obtained previously with Chandra [11]. The X-ray
and gamma-ray constraints thus suggest a population
of about 50 to 60 MSPs in 47 Tuc. This is a factor of
$\sim$2 above the number of known radio MSPs in the
cluster and also well above the upper limit of 30 radio
MSPs estimated to be present in 47 Tuc [27], constraining
the radio beaming fraction to $>$0.5 times that
of the gamma-ray beaming.  We recall that this result
relies on an estimate of the average gamma-ray efficiency
of MSPs. We obtained this efficiency from a
sample of Galactic field MSPs and selected the value
that appears to be the least biased and that is the most
conservative in the sense that it produces the largest
estimate for the number of MSPs in 47 Tuc. The
smallness of the gamma-ray sample of MSPs, however,
implies that the average gamma-ray efficiency
is still uncertain and likely biased.

It has been suggested that MSPs may produce
relativistic magnetized winds that, when interacting
with stellar winds or winds from other MSPs, create
shocks that are capable of accelerating electrons and
positrons into the GeV-TeV regime [28, 29]. These
high-energy particles may eventually undergo inverse
Compton scattering on stellar and cosmic microwave
background radiation, producing detectable
fluxes of GeV-TeV gamma-ray emissions. The expected
gamma-ray emission from 47 Tuc has been
modeled in this scenario by [28], who predicted 100 MeV to 10 GeV photon
fluxes of the order of 10$^8$ photons cm$^{-2}$ s$^{-1}$ and energy
fluxes of the order of $\sim$3 $\times$ 10$^{-11}$ ergs cm$^{-2}$ s$^{-1}$,
which are on the order of the values we observed
using the LAT. However, [28]  assumed
for their calculations that the total power injected
as relativistic electrons and positrons amounts
to 1.2 $\times$ 10$^{35}$ ergs s$^{-1}$, which corresponds to a mean
MSP spin-down power of $<$\.E$>$ = 5.2 $\times$ 10$^{35}$ ergs s$^{-1}$
under the assumptions that the total number of
MSPs in 47 Tuc amounts to 23 objects and that the
average energy conversion efficiency from the pulsar
winds to relativistic electrons and positrons amounts
to 1\% [24, 28]. This spin-down power is about a
factor of 30 larger than the average spin-down
luminosity of MSPs in 47 Tuc, which suggests
that the gamma-ray flux estimates of [28] are overly
optimistic and that the contribution of pulsar wind
interactions to the gamma-ray emission observed by
LAT is probably negligible.

Furthermore, the pulsar wind interaction model
of [28] predicts gamma-ray spectra that extend well
above 1 GeV into the TeV domain, with possible
spectral turnovers and breaks above $\sim$100 GeV.
These high cutoff energies are at odds with our observed
spectral break energy in the GeV range. To
explain a GeV spectral break in the pulsar wind interaction
model, the maximum energy of the accelerated
particles should be limited to a few GeV;
this scale is below the injection energies expected for
electrons and positrons from the inner pulsar magnetospheres,
which may range up to TeV energies
[29]. Consequently, pulsar wind interactions should
play a minor role in the acceleration of electrons
and positrons in 47 Tuc. This is consistent with the
model of [24], which suggests that the direct conversion
efficiency of spin-down energy into gamma
rays, $\eta_\gamma$, is considerably larger than the efficiency
$\eta_{e\pm}$ for electron and positron production.  We cannot
exclude, however, the possibility that pulsar wind
interactions contribute at a low level to the gamma-ray
signal we detected from 47 Tuc. Because TeV
gamma-ray emission from 47 Tuc has not yet been
detected [30], we cannot place firm constraints on
that contribution.

Until now the study of close binary systems in
globular clusters has mainly relied on X-ray observations.
Such studies, however, are hampered by the
fact that a large variety of binary systems emit X-rays
[cataclysmic variables (CVs), low-mass x-ray binaries
(LMXBs), chromospherically active main-sequence
binaries (BY Dras/RS CVns), and MSPs] and that it is difficult
to assess the nature of the sources from X-ray observations
alone (however, see [11]). X-ray studies must
therefore be backed up by multiwavelength identification
programs that help to disentangle these
source populations. High-energy gamma-ray observations
are unique in that they should be sensitive
mainly to the pulsar populations. This is illustrated
by Fermi observations of our own Galaxy that have
revealed that pulsars form the largest and most luminous
point-source population in this energy domain.
No CVs, LMXBs, or BY Dras have so far
been detected in high-energy gamma rays [31]. It
thus seems rather likely that pulsars (and MSPs in
particular) are also the primary population of gamma-ray
sources in globular clusters.

\bigskip 
\begin{acknowledgments}
The Fermi LAT Collaboration is supported by NASA and the
U.S. Department of Energy; the Commissariat \`a l'Energie
Atomique and CNRS/Institut National de Physique Nuclaire et
de Physique des Particules (France); the Agenzia Spaziale
Italiana and Istituto Nazionale di Fisica Nucleare (Italy); the
Ministry of Education, Culture, Sports, Science and Technology,
High Energy Accelerator Research Organization (KEK), and
Japan Aerospace Exploration Agency (Japan); and the
K. A. Wallenberg Foundation, Swedish Research Council, and
National Space Board (Sweden). Additional support was
provided by the Istituto Nazionale di Astrofisica (Italy) and the
Centre National Etudes Spatiales (France).

\end{acknowledgments}

\bigskip 

\end{document}